\documentclass[a4paper,11pt]{article}

\usepackage{amsmath,amsfonts,amsthm}
\usepackage{yfonts}
\usepackage{bbm}
\usepackage{bm}
\usepackage{mathrsfs}

\usepackage{graphicx,color}
\usepackage[T1]{fontenc}
\usepackage{cases}

\newcommand{\lk}{\left(}
\newcommand{\rk}{\right)}
\newcommand{\lkk}{\left\{ }
\newcommand{\rkk}{\right\} }

\newcommand{\be}{\begin{equation}}
\newcommand{\ee}{\end{equation}}
\newcommand{\e}{{\rm e}}
\newcommand{\mD}{{\rm{D}}}

\newcommand{\mS}{\mathcal{S}}
\newcommand{\PP}{\mathbb{P}}
\newcommand{\RR}{\mathbb{R}}
\newcommand{\NN}{\mathbb{N}}
\newcommand{\CC}{\mathbb{C}}

\title{Mean first-passage time of a random walker under Galilean transformation}

\author{Marcus Dahlenburg$^{1,2}$ and Gianni Pagnini$^{2,3}$\\
\\
$^1$BCAM -- Basque Center for Applied Mathematics,\\ 
Alameda Mazarredo 14, 48009 Bilbao, Basque Country -- Spain\\
$^2$Institute for Physics \& Astronomy, University of Potsdam,\\
Karl-Liebknecht-St 24/25, 14476 Potsdam, Germany\\
$^3$Ikerbasque -- Basque Foundation for Science,\\
Plaza Euskadi 5, 48009 Bilbao, Basque Country -- Spain
}

\begin{document}
\maketitle

\begin{abstract}
We consider a continuous-time random walk model
with finite-mean waiting-times 
and we study the mean first-passage time (MFPT) as estimated
by an observer in a reference frame $\mS$,
that is co-moving with a target,
and by an observer in a reference frame $\mS'$,
that is in uniform motion with respect to the target
through a Galilean transformation.
We found that the simple picture emerging in $\mS$,
where the mean first-passage time depends 
on the whole jump distribution
but only on the mean value of the waiting-times,
does indeed not hold in $\mS'$
where the estimation depends on the whole jump distribution 
and also on the whole distribution of the waiting-times. 
We derive the class of jump-size
distributions such that the dependence of the MFPT on the mean
waiting-time only is conserved also in $\mS'$. 
However, if the MFPT is finite, 
the dependence on the specific waiting-time distribution disappears in
$\mS'$ when the initial position is sufficiently 
far-away from the target. 
While the MFPT emerges to be Galilean 
invariant with both two-sided and one-sided jump distributions with finite 
moments, the MFPT is not a Galilean invariant for one-sided jump 
distribution with power-law tails (one-sided L\'evy distributions).
\end{abstract}

\section{Introduction}
\label{sec:introduction}
First-passage time statistics is a well established observable 
with an important role in many applications, see,
e.g.,
\cite{bray_etal-ap-2013,
metzler_etal-firstpassagetime-book-2014,
jpaspecialissue-2020},
%searching \cite{} and foraging \cite{} mechanisms,
%reaction phenomena \cite{}, ... \cite{specialissuesujpadiRM} ...,
and one application of special interest concerns 
biophysical jump processes \cite{bressloff_etal-2014} 
and biological systems in general 
\cite{chou_etal-2014,polizzi_etal-ijc-2016}.
While a large number of theoretical results 
are still derived both in the framework of the classical Brownian motion 
\cite{metzler-jsm-2019,
hartich_etal-jpa-2019, 
majumdar_etal-jsm-2020,
grebenkov_etal-njp-2020,
kearney_etal-jpa-2021,
eliazar-jpa-2021}
and in the framework of the anomalous diffusion processes 
within the diffusing-diffusivity approach 
\cite{sposini_etal-jpa-2019,grebenkov_etal-jpa-2021}
or L\'evy-like motions  
\cite{chechkin_etal-jpa-2003,
koren_etal-prl-2007,
koren_etal-pa-2007,
majumdar_etal-jpa-2017,padash_etal-jpa-2019,
padash_etal-jpa-2020,palyulin_etal-njp-2019,nicolau_etal-jpa-2021},
the quantitative experimental analysis of first-passage time distributions
is becoming accessible in biology only recently
\cite{thorneywork_etal-sa-2020,broadwater_etal-bj-2021}
thanks to the increasing availability of single-molecule measurements
\cite{manzo_etal-rpp-2015,miller_etal-rpp-2018}.

At the same time,
foundation of statistical mechanics has been recently tested  
against the principle of Galilean invariance 
\cite{cairoli_etal-pnas-2018} and, actually,
within the Zwanzig approach in the Hamiltonian framework,
the notion of thermal equilibrium emerges to be not frame invariant 
because the required properties of the random force are 
peculiar of a specific reference frame and the noise must acquire
different statistics in a different reference frame
\cite{cairoli_etal-pnas-2018}.
The consequence is that present coarse-grain diffusive models 
meet Galilean invariance in the phase-space only 
- and therefore in a weak sense only - 
because the stochastic forcing represented by the noise is indeed
always stated with the same properties
as independent of any reference frame transformation
\cite{cairoli_etal-pnas-2018}.

Hence, motivated by the (yet) unavoidable failure
of Galilean invariance in stochastic processes \cite{cairoli_etal-pnas-2018} 
and the emerging of experimental investigation 
of first-passage time statistics in biological systems,
we want to study the effect of a Galilean transformation by a mean drift 
\cite{robson_etal-ptrsb-2013,neupane_etal-prl-2018}
on the measurements of the mean first-passage time (MFPT)
with the aim to provide some indications for 
quantifying and eventually for removing this effect.

The continuous-time random walk (CTRW) approach
\cite{kutner_etal-epjb-2017,shlesinger-epjb-2017}
emerged to be successful for understanding,
reproducing and explaining the main features of 
the anomalous diffusion observed in living systems 
\cite{barkai_etal-pt-2012,goiko_etal-bj-2018}. 
In a CTRW model, 
at any iteration occurring after a random waiting-time, 
the walker's displacement is an instantaneous jump and then 
it is independent of any reference frame transformation,
which, with finite-mean waiting-times,  
causes weakly Galilean invariance because the process is local in time;
in opposition to the CTRW with infinite-mean waiting-times
that converges to time-fractional anomalous diffusion
that is nonlocal in time 
and it is not Galilean invariant \cite{cairoli_etal-pnas-2018}.

Here, we consider a one-dimensional CTRW model in continuous space
with finite-mean waiting-times
and asymmetric or one-sided distribution of jumps,
which allows for a {\it finite} MFPT. 
We remind that the MFPT of a CTRW with symmetric jumps 
is infinite both in continuous 
\cite{redner-2001,dahlenburg_etal-jpa-2022} 
and in discrete space with nearest-neighbour jumps 
\cite{gutkowiczkrusin_etal-jsp-1978,
balakrishnan_etal-p-1983,berezhkovskii_etal-jcp-2008} but, in opposition,
the MFPT of a symmetric diffusion process is finite when 
particle stochastically resets to its initial position at a constant rate
\cite{evans_etal-prl-2011}.
Within this CTRW modelling setting,
we study the estimation of a {\it finite} MFPT
by an observer in a reference frame $\mS$,
that is co-moving with the target,
and by an observer in a reference frame $\mS'$,
that is in uniform motion with respect to the target
according to a Galilean transformation.

The present research can also be re-phrased as a study of 
the MFPT over uniform moving boundaries for a CTRW model,
a similar problem for the CTRW, but with an immobile target only,
was considered by Franke \& Majumdar \cite{franke_etal-jsm-2012},
and the case with a constant time-step but 
general random-moving boundaries 
has been recently analysed for L\'evy flights 
by Denisov {\it et al.} \cite{denisov_etal-tpa-2019}.

We found that in $\mS$
the MFPT results to be dependent on the jump distribution
and on the mean value only of the waiting-times, but 
this simple picture emerged in $\mS$ 
is actually broken in $\mS'$
where the estimation depends on the whole jump distribution
and also on the whole distribution
of the waiting-times. 

Different estimations of the MFPT emerge according to the 
direction of the relative motion between $\mS'$ and the target.
In this setting, 
the averaged relative shift between 
$\mS'$ and the target is provided by  
the frame velocity, the mean waiting-time and the mean jump-size. 
In particular, 
for initial position sufficiently far-away from the target,
the finiteness of the MFPT is determined solely by 
a negative value of the averaged relative shift.
In this limit, 
the two estimations in $\mS$ and $\mS'$ are reconciled 
only with an asymmetric random walk when the jump-size distribution 
is governed by a power-law distribution towards the target, 
which has an infinite mean, 
and an arbitrary jump-size distribution in the direction opposite
to the target, which has a finite mean. 
In this case, the MFPT is not a Galilean invariant. 
We provide also under which conditions the two estimations 
of the MFPT in $\mS$ and $\mS'$ are finite but different, namely, 
when the MFPT meets the Galilean invariance. 
Moreover, we derive the class of jump-size distributions such 
that the dependence of the MFPT on the mean waiting-time only is conserved also 
in $\mS'$, for any initial position, in analogy with $\mS$.
This occurs in the special case of the weighted superposition of 
an exponential jump-size distributions towards a departing target 
together with a non-null probability to rest and 
an arbitrary jump-size distribution with a finite mean in the 
direction opposite to the target.

In the following, we first provide notions and notation 
in next Section \ref{sec:notions} 
and later we derive and discuss
the conditions for a finite MFPT in the reference frame $\mS$
and in the reference frame $\mS'$ 
in Section \ref{mS} and \ref{mSprime}, respectively. 
Conclusions are reported in Section \ref{conclusions}.

\section{Notions and notation}
\label{sec:notions}
Let the two reference frames $\mS$ and $\mS'$ be 
represented by
the one-dimensional coordinate systems $x \in \RR$ and $x' \in \RR$,
respectively, 
and then be related by
the Galilean transformation with constant velocity $v \in \RR$, i.e.,
\be
x' = x + v \, t \,, \quad t \ge 0 \,,
\label{GT}
\ee
where $t$ is the time-lag.
%Moreover, let the target be located in $\mS$ at $x=x_T$.  

A CTRW model for the walker's position 
after $N \in \NN$ iterations, and here denoted by $X_{t_N}$,
is defined in the two reference frames 
$\mS$ and $\mS'$ as follows
\begin{flalign}
{\rm in} \,\, \mS: \quad
X_{t_N}=x_0 + \sum_{i=1}^N \delta x_i \,, \quad
t_N = \sum_{i=1}^N \delta t_i \,, 
\end{flalign}
\be
{\rm in} \,\, \mS': \quad
X'_{t_N}=X_{t_N} + v \, t_N 
= x_0 + v \, t_N + \sum_{i=1}^N \delta x_i \,,
\ee
with initial data
$t_0=0$, $X_0=x_0$, $X'_0=X_0$, and
where $\{\delta x_i\} \in \RR$ and $\{\delta t_i\} \in \RR_0^+$ 
are the jumps and the waiting-times that are independently drawn 
at each iteration from the probability density function (PDF)
$q(\xi)$ and $\psi(\tau)$, respectively.
In general, density $q(\xi)$ can be also asymmetrical.
Then, peculiar characteristic of the CTRW is that at the 
observation time $t=t_N$ each realisation of the 
walker's trajectory has performed a corresponding
number $N$ of jumps.
Moreover, let the symbols $\PP$ and $\PP'$ be the probability measure
in the frame $\mS$ and $\mS'$, respectively, then it holds
\be
\PP(X_t \le x) = 
\PP'(X'_t \le x') \,.
\label{identityPP}
\ee
By definition we have that 
\be
\PP(X_t \le x) = 
\int_{-\infty}^{x} P(z;t) \, dz \,,
\label{defPP}
\ee
where $P(x;t)$ is the PDF estimated in $\mS$ 
for a walker to be in $x$ at time $t$, such that
$\displaystyle{\int_{-\infty}^{+\infty} P(x;t) dx = 1}$ and $P(x;t) > 0$
for all $(x,t) \in \RR \times (0,+\infty)$ 
with initial datum $P(x;0)=P_0(x)$.
The same definition holds for $\PP'$ in $\mS'$:
\be
\PP'(X'_t \le x') =
\int_{-\infty}^{x'} P'(y;t) \, dy \,.
\label{defPPprime}
\ee
Hence, equality (\ref{identityPP}) reads
\be
\int_{-\infty}^{x} P(z;t) \, dz =
\int_{-\infty}^{x'} P'(y;t) \, dy \,.
\label{equationP}
\ee

Equality (\ref{equationP}) is important for
establishing the relation between PDFs in $\mS$ and $\mS'$ 
and then for the comparison between MFPT estimations
presented in the following.
In particular, 
by applying transformation (\ref{GT}) to (\ref{equationP}),
it emerges that
\begin{eqnarray}
\int_{-\infty}^{x} P(z;t) \, dz 
&=& \int_{-\infty}^{x'} P'(y;t) \, dy \nonumber \\
&=& \int_{-\infty}^{x+vt} P'(y;t) \, dy 
= \int_{-\infty}^{x} P'(z+vt;t) \, dz \,, 
\end{eqnarray}
and then 
\be
P(x;t)=P'(x+vt;t) \,\, \text{and} \,\, P(x-vt;t)=P'(x;t) \,,
\label{identity}
\ee
that takes into account Galilean transformation (\ref{GT}) 
for particles' density.

%\textcolor{red}{
%First passage time statistics as a physical observable
%dated back
%is a well established can be referred to
%as a dated problem \cite{SmoluchowskiMV1917Z.Phys.Chem.92129
%[5] CollinsFCandKimballGE1949J.ColloidSci.4425}, but 
%QUI parlare della importanza del MFPT e del CTRW nella letteratura.
%}

\section{Analysis in the co-moving reference frame $\mS$}
\label{mS}
\subsection{Derivation of the MFPT formula}
Within the CTRW approach for particle diffusion,
the determination of the walker's PDF
is a two-step procedure \cite{zaburdaev_etal-rmp-2015}
that in $\mS$ is given by the formulae
\begin{subequations}
\be
\Omega(x;t) = P_0(x)\psi(t) + \int_0^t \psi(t-\tau) 
\int_{-\infty}^{+\infty} q(x-\xi)
\Omega(\xi;\tau) \, d\xi d\tau \,,
\label{equationomega}
\ee
\be
P(x;t) = P_0(x)\Psi(t) + \int_0^t \Psi(t-\tau) 
\int_{-\infty}^{+\infty} q(x-\xi)
\Omega(\xi;\tau) \, d\xi d\tau \,,
\label{equationp}
\ee
\end{subequations}
where $\Omega(x;t)$ is the outgoing normalised flow of 
particles that counts the amount of walkers are leaving  
the point $x$ per unit of time at the instant $t$,
and $\displaystyle{\Psi(t)=1-\int_0^t \psi(\tau) d\tau}$ is 
the probability for a walker to stay in the same location
until time $t$.

By applying Fourier and Laplace transformations to 
(\ref{equationomega}) and to (\ref{equationp}), and 
by solving with respect to $P(x;t)$, we obtain
\be
\widehat{\widetilde{P}}(\kappa;\lambda)=
\frac{\widehat{P}_0(\kappa) \widetilde{\Psi}(\lambda)}
{1 - \widehat{q}(\kappa) \widetilde{\psi}(\lambda)} \,,
\label{MWeq}
\ee
that is the Montroll--Weiss equation 
\cite{zaburdaev_etal-rmp-2015, montroll_etal-jmp-1965},
where $\kappa \in \RR$ is the wavenumber parameter
in the Fourier domain and
$\lambda \in \CC$ is the frequency parameter in the Laplace domain.
In the physical domain, it holds
\be
P(x;t) = \Psi(t)P_0(x) + \int_0^t \psi(t-\tau) 
\int_{-\infty}^{+\infty} q(x-\xi)
P(\xi;\tau) \, d\xi d\tau \,.
\label{eq1}
\ee

The walker's PDF $P(x;t)$ can be stated in terms of the propagator, namely
of the conditional probability density $p(x;t|x_0)$, i.e., 
\be
P(x;t)=\int_{-\infty}^{+\infty} p(x;t|x_0) P_0(x_0) \, dx_0 \,,
\label{eq2}
\ee
and
%\textcolor{red}{[this is not necessary: when $P_0(x)=\delta(x-x_0)$]} 
from (\ref{eq1}) we obtain
\be
p(x;t|x_0) = \Psi(t)\delta(x-x_0) + \int_0^t \psi(t-\tau) 
\int_{-\infty}^{+\infty} q(x-\xi) p(\xi;\tau|x_0) \, d\xi d\tau \,,
\label{eq3}
\ee
where spatial homogeneity of the process has been assumed, 
i.e., $p(x;t|x_0)=p(x-x_0;t|0)$.

In order to estimate the MFPT for the observer in $\mS$, 
we locate an absorbing boundary at the target position $x=x_T$ 
and we compute the survival probability of the walker,
i.e., the probability for the walker to be not beyond the target. 
Without loss in generality, 
since the process is assumed spatially homogeneous,
%and diffusion occurs in an unbounded domain, 
we set $x_0 > x_T = 0$ and we look for the equation
governing the walker's PDF laying in $x \ge 0$. 
In particular, we stress that we analyse the problem at the level
of the equation and not at the level of the solution, 
thus the method of imagines - or similar - does not apply
but proper initial and boundary conditions are needed.
In our notation, due to the absorbing events, 
we look for the propagator provided only by the non-absorbed particles
that means provided by those walker's realisations
that at the previous iteration were in the positive semi-axis by taking
advantage of the renewal nature of the CTRW. 
With reference to the integral over $\xi$ in (\ref{eq3}),
by using the properties of convolution integrals, 
we first replace the integrand $q(x-\xi)p(\xi;\tau|x_0)$ with
$q(\xi)p(x-\xi;\tau|x_0)$ and
later we introduce the jump-size with respect to the starting point
by the shifting $\xi \to \xi-x_0$ such that the integrand
turns out to be $q(\xi-x_0)p(x;\tau|\xi)$.
Moreover,
in analogy with a killing process,
we remove the contribution from the sourcing points $\xi$ 
that are beyond the target by taking the integral interval  
over the positive semi-axis only. In formulae:
\begin{eqnarray}
p_\text{abs}(x;t|x_0)
&=& \Psi(t) p_\text{abs}(x;0|x_0) + \nonumber \\
& & \int_{0}^t \psi(t-\tau)\int_{0}^\infty 
q(\xi - x_0)p_\text{abs}(x;\tau|\xi) \, d\xi d\tau \,, 
\label{eq4}
\end{eqnarray}
with $(x,t) \in [0,\infty)\times(0,\infty)$,
$p_\text{abs}(x;0|x_0)=\delta(x-x_0)$ and 
$p_\text{abs}(0;t|x_0)=0$.
The effect of the absorbing boundary is embodied by the non-normalisation
of $p_\text{abs}(x;t|x_0)$. 

%\textcolor{red}{
%THIS IS CORRECT BUT UNNECCESARY - 
%In fact, by applying the Laplace transform to (\ref{eq4}) 
%both in $x$ and $t$ with Laplace parameters $s$ and $\lambda$ and symbols 
%$\, \widetilde{} \,$ and $^*$, respectively, we have   
%\begin{eqnarray}
%\widetilde{p}^*_\text{abs}(\lambda;s|x_0) &=& 
%\frac{\e^{-\lambda x_0}}{s} -
%\nonumber \\
%& & \hspace{-1.9truecm}
%\widetilde{\psi}(s)\left[\frac{\e^{-\lambda x_0}}{s} -
%\int_0^\infty q(\xi - x_0) \widetilde{p}^*_\text{abs}(\lambda;s|\xi) 
%\, d\xi \right] \,,
%\end{eqnarray}
%from which it emerges that the normalization of 
%$p_\text{abs}(x;t|x_0)$ in (\ref{eq4}),  
%namely $\widetilde{p}^*_\text{abs}(0;s|x_0) = 1/s$, 
%implies the condition
%$\displaystyle{\int_0^\infty q(\xi - x_0) \, d\xi=
%\int_{-x_0}^\infty q(\xi) \, d\xi=1}$,
%but it is in opposition
%to the assumed spatial homogeneity of $q(\xi)$. 
%Therefore, function $p_\text{abs}(x;t|x_0)$ in (\ref{eq4}) is not
%normalized.
%}

Therefore, 
the survival probability - hereinafter denoted by $\Lambda(x_0,t)$ -
results to be 
\begin{eqnarray}
\Lambda(x_0,t)
&=& \int_0^\infty p_\text{abs}(x;t|x_0) \, dx \nonumber \\
&=& \Psi(t)+\int_{0}^t \! \psi(t-\tau) \! \int_{0}^\infty  
\!\! q(\xi - x_0) \Lambda(\xi,\tau) \, d\xi d\tau \,,
\label{eq5}
\end{eqnarray}
that meets the conditions $\Lambda(x_0,0)=1$ and
$\lim_{t\to+\infty}\Lambda(x_0,t)=0$. 
We highlight that if the waiting-times are all equal and not distributed, 
namely $\psi(\tau)=\delta(\tau - \Delta t)$,
with $\Delta t > 0$, and 
$t=n \Delta t$, with $n \in \NN$, such that
$\Psi(0)=1$ and
$\displaystyle{
\Psi(t \ge \Delta t)=1-\int_0^t \delta(\tau-\Delta t) \, d\tau=0}$,
then from (\ref{eq5}) 
the equation for the survival probability with discrete time
is recovered, see, e.g., \cite[equation (10)]{bray_etal-ap-2013}
and \cite[equation (18)]{dahlenburg_etal-jpa-2022}.
In particular, in the discrete-time case,
the survival probability is governed by
a homogeneous Wiener--Hopf equation.
% while, in the continuous-time case, 
% by an nonhomogeneous Wiener--Hopf equation.   
Moreover, we stress here that the results by 
Sparre Andersen \cite{sparre-andersen-1953,sparre-andersen-1954b} 
as well as those by
Pollaczek \cite{pollaczek-1952} and Spitzer 
\cite{spitzer-1956,spitzer-1957,spitzer-1960} 
for the determination of the survival probability
$\Lambda(x_0,t)$ hold for Markovian random walks with discrete time
and that the CTRW  
%beside a distribution of the waiting-times between two consecutive jumps
%that leads to an nonhomogeneous rather than homogeneous
%Wiener--Hopf equation, 
is indeed Markovian if and only if the 
waiting-times distribution is exponential 
\cite{zwanzig-jsp-1983,mainardi_etal-pa-2000}.

The PDF of the FPT is defined as
$\displaystyle{
\mathcal{P}_{x_0}(t)=-\frac{\partial \Lambda(x_0,t)}{\partial t}}$
and then the MFPT - hereinafter denoted by $T(x_0)$ -  
is here estimated by the following limit 
\begin{eqnarray}
T(x_0) &=& 
\int_0^\infty t \, \mathcal{P}_{x_0}(t) \, dt 
= - \int_0^\infty t \, \frac{\partial \Lambda(x_0,t)}{\partial t} \, dt 
\nonumber \\
&=& \int_0^\infty \Lambda(x_0,t) \, dt =
\lim_{\lambda \to 0} \widetilde{\Lambda} (x_0,\lambda) \,,
\label{defLambda}
\end{eqnarray}
where $\widetilde{\Lambda} (x_0,\lambda)$ is the Laplace transform of 
the survival probability that from (\ref{eq5}) is determined by the
equation
\begin{equation}\label{eq6}
\widetilde{\Lambda}(x_0,\lambda)
=\widetilde{\Psi}(\lambda)+ \widetilde{\psi}(\lambda)
\int_{0}^\infty q(\xi - x_0) \widetilde{\Lambda}(\xi,\lambda) \, d\xi \,.
\end{equation}
By definitions, it holds 
$\widetilde{\Psi}(\lambda)=(1-\widetilde{\psi}(\lambda))/\lambda$
and $\widetilde{\psi}(0)=1$ so, 
by assuming a finite-mean waiting-times distribution, i.e., 
$\displaystyle{
\langle \tau \rangle = \int_0^\infty \tau \, \psi(\tau) \, d\tau < \infty}$,
we have that $\widetilde{\psi}(\lambda) \sim 
1 - \langle \tau \rangle \lambda + o(\lambda)$
and then the MFPT (\ref{defLambda})
estimated by an observer in co-motion with
the target is
\be
T(x_0)=\langle \tau \rangle + \int_{0}^\infty  q(\xi - x_0)T(\xi) \, d\xi 
\,, \quad x_0 > 0 \,.
\label{eq7}
\ee
In particular, from (\ref{eq7}), 
we observe that the MFPT as estimated in $\mS$ depends 
on the full distribution of the jumps' size 
but on the mean value only of the waiting-times, 
actually, $T(x_0)$ is independent of the specific form of the PDF
of the waiting-times $\psi(\tau)$. 

Formula (\ref{eq7}) is an non-homogeneous Wiener--Hopf integral equation
that admits solution $T(x_0)=\mathcal{O}(\e^{b x_0})$ as $x_0 \to \infty$
whenever $q(\xi)=\mathcal{O}(\e^{-a|\xi|})$ as $\xi \to \infty$ with
$a > b$ given that $\langle \tau \rangle = \mathcal{O}(\e^{b x_0})$
when $x_0 \to \infty$, see \cite[p. 345]{integralequations}.
Nevertheless, not all solutions are also MFPT solutions
\cite{dahlenburg_etal-jpa-2022}. 

\subsection{Two-sided jumps} 
By remembering that the MFPT for symmetric diffusive processes
is infinite,
a finite MFPT solution can be obtained with 
a two-sided asymmetric jump-size distribution with exponential tails that,
by taking into account the above conditions on the existence of the 
solution of (\ref{eq7}), is indeed quite a general case. 
In particular, such MFPT solutions can be explicitly derived by taking 
\cite{dahlenburg_etal-jpa-2022}
\be
q(\xi)=\left\{
\begin{array}{ll}
(1-b) \exp(\xi) \,, & \xi \in \RR^-_0 \,, \\
\\
ab \exp(-a\xi) \,, & \xi \in \RR^+ \,,
\end{array}
\right.
\label{expPDF}
\ee
with $b \in [0,1]$ and $a \in \RR^+$, where the mean value is
\be
\langle \xi \rangle = \left(\frac{1+a}{a}\right) b - 1 \,,
\label{expMEAN}
\ee
and it holds \cite{dahlenburg_etal-jpa-2022}
\be
T(x_0)=-\frac{\langle \tau \rangle}{\langle \xi \rangle}(1+x_0) \,, \quad
x_0 \,, \langle \tau \rangle \,, - \langle \xi \rangle \in \RR^+ \,.
\label{MFPT-WH}
\ee
We highlight that the MFPT is infinite in some asymmetric cases, too,
whenever $\langle \xi \rangle = 0$ which corresponds to
$b=a/(1+a)$.
Formula (\ref{MFPT-WH}) in the asymptotic limit 
$x_0 \gg 1$ reduces to the known result 
\cite{redner-2001}, 
\be
T(x_0) \sim - \frac{x_0}{u} \,, \quad 
u=\frac{\langle \xi \rangle}{\langle \tau \rangle} < 0 \,,
\label{asymptotic}
\ee 
for a starting point far-away enough from the boundary. 

We report that the emergence of Wiener--Hopf equations 
in boundary problems for diffusive processes is,
as a matter of fact, the proper mathematical scenario, see, e.g., 
\cite{bray_etal-ap-2013,
majumdar_etal-jpa-2017,
spitzer-1957,spitzer-1960,
ivanov-aa-1994,
frisch_etal-1994,
majumdar_etal-jsp-2006,
majumdar-pa-2010}. 

The non-universality of the MFPT for starting points near
the boundary has been recently discussed by these authors
\cite{dahlenburg_etal-jpa-2022}.
In fact, if we pass to the spatial-dimensional setting through the length-scale
$\ell$ we have  
$x^{\mD}_0 = \ell \, x_0$ and $\xi^{\mD}=\ell \, \xi$ such that 
$\langle \xi^{\mD} \rangle = \ell \, \langle \xi \rangle$ 
from which it follows $\ell=\langle \xi_\mD \rangle / \langle \xi \rangle$.
Hence, even if the spatial-dimensional processes have the same
mean jump-size $\langle \xi_\mD \rangle$, 
the length-scale of the systems $\ell$ can be in general different.
In fact, $\ell$ depends also on $\langle \xi \rangle$ (\ref{expMEAN}) 
that is a function of the specific values of parameters $a$ and $b$ 
building-up the jump-size distribution (\ref{expPDF}). 
More quantitatively, in the spatial-dimensional setting, 
formula (\ref{MFPT-WH}) reads 
\be
T(x_0)=
-\frac{\langle \tau \rangle}{\langle \xi \rangle}
\left[ 1  + \frac{x_0^{\mD}}{\ell} \right] =
-\frac{\langle \tau \rangle}{\langle \xi \rangle}
\left[ 1  + \langle \xi \rangle 
\frac{x_0^{\mD}}{\langle \xi^{\mD} \rangle} \right] \,,
\ee
with $x_0^{\mD}$, $\langle \tau \rangle$, $- \langle \xi \rangle \in \RR^+$,
from which it emerges that the universal asymptotic limit (\ref{asymptotic}), 
that in spatial-dimensional form is
\be
T(x_0) \sim 
-\frac{\langle \tau \rangle}{\langle \xi^{\mD} \rangle}
\, x_0^{\mD} \,,
\ee
is obtained when
\be
x_0^{\mD} \gg \frac{\langle \xi^{\mD} \rangle}{\langle \xi \rangle} \,,
\label{nonuniversal}
\ee
that is non-universal in spite of equal spatial-dimensional mean jump-size
$\langle \xi^{\mD} \rangle$ because $\langle \xi \rangle$
depends on the values of $a$ and $b$ that are proper of each 
jump-size PDF (\ref{expPDF}).

Summarising, 
the length-scale discriminating 
between starting-points near and far the boundary is not universal.
The family of asymmetric exponential-distributions 
that we have studied (\ref{expPDF}) defines a common length-scale 
for a sub-family of PDFs with the same mean (\ref{expMEAN}) 
for different values of parameters $a$ and $b$ 
after which the MFPT is universal for all the processes of the sub-family, 
while the MFPT is not universal for starting-points less than such length-scale.

To conclude, 
from formula (\ref{MFPT-WH}) it emerges that
when the jump-PDFs $q(\xi)$ are two-sided with exponential tails 
then the MFPT is finite if $\langle \xi \rangle < 0$, 
i.e., in the direction of the target. 
Thus the asymmetry of the processes does not guarantee a finite MFPT
because those with $\langle \xi \rangle \ge 0$ lead indeed
to an infinite MFPT together with symmetric processes.

Similarly, also in the case of 
two-sided jump-PDFs with power-law tails 
the asymmetry of the processes does not guarantee a finite MFPT.
In fact, when the resulting random walk is a L\'evy flight
with stable index
$\alpha \in (0,1) \cup (1,2)$ and asymmetry parameter
$\beta \in (-1,1)$ then the MFPT is finite when
$\alpha \in (1,2)$ 
\cite{padash_etal-jpa-2019,padash_etal-jpa-2020}.
See in Appendix \ref{AppendixA} its determination.
In the other asymmetric case, namely $\alpha \in (0,1)$,
together with the symmetric ones, the MFPT is infinite
\cite{
chechkin_etal-jpa-2003,
koren_etal-prl-2007,
koren_etal-pa-2007,
majumdar_etal-jpa-2017,padash_etal-jpa-2019,
padash_etal-jpa-2020,palyulin_etal-njp-2019}.
It is important to report that 
without an absorbing boundary the PDF decreases according
to the following limit behaviours on the left of the starting point:
when $\alpha \in (1,2)$ and $\beta=1$ the PDF is two-sided
and it has a right power-law tail $x^{-1-\alpha}$ when $x \to + \infty$ and 
a left exponential tail lighter than the Gaussian when $x \to - \infty$.
In these cases the MFPT is finite with an
absorbing boundary at $x_T < x_0$. 
When $\beta=-1$ the properties of the tails are exchanged
and then the MFPT is finite when the absorbing boundary is at
$x_T > x_0$.

\subsection{One-sided jumps}
\label{onesideS}
We consider now one-sided jumps in the direction of the target
and the MFPT emerges to be finite both with a jump PDF
with finite moments and with a power-law jump PDF.
Within this framework, we have
\be
\int_{-\infty}^0 q(\xi) \, d\xi=1 \,,
\quad q(\xi) > 0 \quad {\rm for} \,\, {\rm all} \quad \xi \in \RR^-_0 \,,
\label{eq8}
\ee
and the MFPT results to be governed by the following 
Volterra equation of the second kind 
\be
T(x_0)
= \langle \tau \rangle +\int_{0}^{x_0}  q(\xi - x_0)T(\xi) \, d\xi \,,
\label{Teq}
\ee
rather than by the non-homogeneous Wiener--Hopf equation (\ref{eq7}).
From (\ref{Teq}) we have that
\be
\widetilde{T}(s) = \frac{\langle \tau \rangle}
{s \, [1-\widetilde{\rho}(s)]} \,,
\label{eq10}
\ee
where
%where $q^{+}(-\xi) = q(\xi)$, with $\xi \in \RR^-_0$, such that 
%with the change of variable $-\xi \to \theta \in \RR^+_0$ 
%we have $\displaystyle{\widetilde{q}^{+}(s)=\int_0^\infty 
%\e^{-s\theta} \, q^{+}(\theta) \, d\theta}$. 
\be
\widetilde{\rho}(s)=\int_0^\infty 
\e^{-s\theta} \, \rho(\theta) \, d\theta \,,
\quad \text{where} \quad \rho(\theta)=\rho(-\xi) = q(\xi) \,,
\label{theta}
\ee
with $-\xi = \theta \in \RR^+_0$.
Hence, since it holds $\widetilde{\rho}(s) < 1$ when $s \ne 0$, 
then formula (\ref{eq10}) can be re-written as
\be
\widetilde{T}(s) = 
\frac{\langle \tau \rangle}{s} 
\sum_{n=0}^\infty \left[ \widetilde{\rho}(s) \right]^n \,,
\ee
such that by anti-transformation we have
\be
T(x_0) = \langle \tau \rangle
\sum_{n=0}^\infty \int_0^{x_0} \rho_{n}(\theta) \, d\theta \,,
\label{eq11}
\ee
where
$\displaystyle{\rho_{n}(\theta)=
\int_0^\theta \rho_{n-1}(\chi) \rho(\theta-\chi) \, d\chi}$ 
with $\rho_{0}(\theta)=\delta(\theta)$.
Formula (\ref{eq11}) is the unique solution of (\ref{Teq})
and it meets all the constraints for actually being a MFPT,
in opposition with the two-sided case governed by (\ref{eq7})
where solutions not consistent with a MFPT can emerge
\cite[Appendix A]{dahlenburg_etal-jpa-2022}.

If we consider a jump PDF with existing moments such that 
in the limit $s \to 0$ it holds 
$\widetilde{\rho}(s) \sim 1 + \langle \xi \rangle s$,
with $\langle \xi \rangle < 0$, then from (\ref{eq10}) we obtain
\be
T(x_0) \sim - \frac{\langle \tau \rangle}{\langle \xi \rangle} \, x_0 \,, 
\quad x_0 \to \infty \,.
\label{TonesideS}
\ee
We observe that also in this case with one-sided jump-PDFs 
the MFPT depends on the mean value only of the waiting times
but on the whole distribution of the jump-sizes because 
when we move to the dimensional case the discussion around formula 
(\ref{nonuniversal}) still applies.

If we consider a one-sided L\'evy distribution for the jumps then 
$\displaystyle{\widetilde{\rho}(s)=\e^{-s^\alpha}}$, 
with $0 < \alpha < 1$, and from (\ref{eq10}) it holds
\be
\displaystyle{
\widetilde{T}(s)=\frac{\langle \tau \rangle}{s \, [1-\e^{-s^\alpha}]} \,,
}
\ee
that in the limit $s \to 0$ gives
\be
\widetilde{T}(s) \sim \frac{\langle \tau \rangle}{s^{1+\alpha}} \,,
\quad s \to 0 \,,
\ee
and then in the asymptotic case $x_0 \to +\infty$ it results
\be
T(x_0) \sim \frac{\langle \tau \rangle}{\Gamma(1+\alpha)} \, x_0^\alpha \,,
\quad 0 < \alpha < 1 \,,
\quad x_0 \to +\infty \,,
\label{MFPTonesideS}
\ee
that is consistent with literature results, inter alia,
\cite[formula (56)] {padash_etal-jpa-2020}. 

\section{Analysis in the uniform moving reference frame $\mS'$}
\label{mSprime}
\subsection{Derivation of the MFPT formula}
We study now as the same process is described 
by an observer in $\mS'$. 
Actually, by using the identity (\ref{identity})
for deriving formulae (\ref{equationomega}) and (\ref{equationp})
for $\Omega'(x;t)$ and $P'(x;t)$, 
the corresponding Montroll--Weiss equation (\ref{MWeq}) results to be 
\be
\widehat{\widetilde{P'}}(\kappa;\lambda)=
\frac{\widehat{P}_0(\kappa)\widetilde{\Psi}(\lambda - i \kappa v)}
{1 - \widehat{q}(\kappa)\widetilde{\psi}(\lambda - i\kappa v)} \,,
\label{MW_moving}
\ee
see also Reference \cite[equation (12)]{dahlenburg_etal-pre-2021},
and in the physical space it holds 
\be
P'(x;t)
= \Psi(t)P_0(x-vt) + \int_0^t \psi(t-\tau)
\int_{-\infty}^{+\infty} q(x-\xi-v\tau)
 P'(\xi;\tau) \, d\xi d\tau \,.
\label{equationP'}
\ee
By applying again formula (\ref{eq2}) 
in the reference frame $\mS'$ 
%\textcolor{red}{THIS IS NOT NECESSARY: with initial datum 
%$P'_0(x')=P'_0(x)=P_0(x)=\delta(x-x_0)$,}
where identity (\ref{identity}) has been used,
from formula (\ref{equationP'})  
the conditional probability density $p'(x;t|x_0)$ is  
determined by
\begin{eqnarray}
p'(x,t|x_0)
&=& \Psi(t)\delta(x-x_0-vt) + \nonumber \\
& & \qquad \qquad \int_0^t \psi(t-\tau) \int_{-\infty}^{+\infty} 
q(x-\xi-v\tau) p'(\xi;\tau|x_0) \, d\xi d\tau \,, \nonumber\\
&=& \Psi(t)\delta(x-x_0-vt) + \nonumber \\
& & \qquad \int_0^t \psi(t-\tau)
\int_{-\infty}^{+\infty} 
q(\xi-x_0-v\tau) p'(x;\tau|\xi) \, d\xi d\tau \,. %\nonumber\\
%&=&\Psi(t)\delta(x-x_0-vt) + \nonumber \\
%& & \qquad \int_0^t \psi(t-\tau)
%\int_{-\infty}^{+\infty} \delta(y-x_0-v\tau)
%\int_{-\infty}^{+\infty} q(\xi-y) p'(x;\tau|\xi) \, d\xi dy d\tau 
\label{equationpprime}
\end{eqnarray}

When an absorbing boundary is located at $x_T=0$, 
in analogy with (\ref{eq4}) the walker's PDF from
(\ref{equationpprime}) turns into 
\begin{eqnarray}
\hspace{-2.0truecm} 
p'_\text{abs}(x;t|x_0)
%&=& \Psi(t)\Theta(x_0+vt)\delta(x-x_0-vt) + 
%\int_{0}^t \psi(\tau) 
%\int_{0}^\infty \delta(y-x_0-v\tau)
%\int_{0}^{\infty} q(\xi-y)p'_{\text{abs}}(x;t-\tau|\xi) \, d\xi dy d\tau \,,
%\nonumber \\
&=&\Psi(t)\Theta(x_0+vt)\delta(x-x_0-vt) + \nonumber \\
& & \int_{0}^t \psi(t-\tau) \Theta(x_0+v\tau)
\int_{0}^{\infty} q(\xi-x_0-v\tau)p'_{\text{abs}}(x;\tau|\xi) \, d\xi d\tau 
\,,
\label{eq15}
\end{eqnarray}
where again we have removed the contribution from the sourcing points 
$\xi$ that are beyond the target by taking the integration interval 
over the positive semi-axis only and
we have removed also the contribution from those walker's realisation
that did not jump yet but they moved beyond the target 
because of the frame translation
by employing the function $\displaystyle{\Theta(x)=\frac{d}{dx}\max\{0,x\}}$.
Therefore the corresponding survival probability is
\begin{eqnarray}
\hspace{-2.0truecm} 
\Lambda'(x_0,t)
&=& \Psi(t) \Theta(x_0+vt) + \nonumber \\
& & \quad
\int_0^t \psi(t-\tau) \Theta(x_0+v\tau)
\int_{0}^{\infty} q \lk \xi-x_0-v\tau \rk \Lambda(\xi,\tau) 
\, d\xi d\tau \,,
\label{eq16}
\end{eqnarray}
and its Laplace transform 
\begin{eqnarray}
\hspace{-2.5truecm} 
\widetilde{\Lambda}'(x_0,\lambda)
&=&\int_0^\infty \exp(-\lambda t)\Psi(t)\Theta(x_0+vt)dt + \nonumber \\
& &\int_0^\infty \exp(-\lambda\tau) \psi(\tau) \Theta(x_0+v\tau)\int_0^{\infty} 
\!\!
q\lk \xi-x_0-v\tau\rk \widetilde{\Lambda}'(\xi,\lambda) \, d\xi d\tau \,, 
\label{eq17}
\end{eqnarray}
such that, finally, the MFPT estimated by an observer in 
uniform motion with respect to the target is
\begin{eqnarray}
T'(x_0) &=& 
\int_0^\infty 
\Theta(x_0+vt) \Psi(t) dt + \nonumber \\
& & \int_0^\infty \psi(\tau) \Theta(x_0+v\tau)
\int_0^{\infty} q(\xi-x_0-v\tau) T'(\xi) \, d\xi d\tau \,.
\label{eq18}
\end{eqnarray}
From equation (\ref{eq18}), 
we observe that in $\mS'$ 
the MFPT is no more dependent only 
on the mean value of the waiting-times as in $\mS$, see formula (\ref{eq7}), 
but it depends on its whole distribution $\psi(\tau)$
and the related $\Psi(\tau)$.
However, when $v=0$ we recover formula (\ref{eq7}) as expected.

In particular, we distinguish the cases $v < 0$ and $v \ge 0$
that we denote $T'_<(x_0)$ and $T'_>(x_0)$, respectively,
such that (\ref{eq18}) turns into
\be
T'_{<}(x_0) = 
\int_0^{x_0/|v|} \hspace{-0.3truecm} \Psi(\tau) d\tau + 
\int_0^{x_0/|v|} \hspace{-0.3truecm} \psi(\tau) 
\int_0^{\infty} \hspace{-0.2truecm} q(\xi-x_0 - v\tau) 
T'_{<}(\xi) \, d\xi d\tau \,,
\quad v<0 \,, 
\label{MFPT_negative_v}
\ee
\be
T'_{>}(x_0) = \langle\tau\rangle + 
\int_0^{\infty} \psi(\tau) 
\int_0^{\infty} q(\xi-x_0-v\tau) T'_{>}(\xi) \, d\xi d\tau \,,
\quad v\ge 0 \,.
\label{MFPT_positive_v}
\ee
Since it holds 
\be
\int_0^{x_0/|v|} \Psi(\tau) \, d\tau < 
\int_0^\infty \Psi(\tau) \, d\tau = \langle \tau \rangle \,,
\ee
and also
\be
\int_0^{x_0/|v|} \,\, \psi(\tau) \, d\tau <
\int_0^\infty \psi(\tau) \, d\tau = 1 \,,
\ee
it results that 
\be
T'_<(x_0) < T'_>(x_0) \,, \quad v \neq 0 \,.
\ee

\subsection{Two-sided jumps}
We consider also in this case the exponential PDF defined in (\ref{expPDF}).
A solution of equation (\ref{MFPT_negative_v}) can be obtained
and we have the explicit result
\begin{eqnarray}
\widetilde{T}'_{<}(s) &=&
\left[
\frac{\widetilde\Psi(|v| s)}{s}-
\frac{ab}{a-s}\widetilde{\psi}(|v| s)\widetilde{T}'_{<}(a)\right]
\nonumber \\
& & \times
\frac{(a-s)(1+s)}{(a-s)(1+s)-\widetilde{\psi}(|v| s)[(1-b)(a-s)+ ab(1+s)]} \,,
\end{eqnarray} 
where we observe that the MFPT depends on the whole distribution
of the waiting-times, and not on the mean value only,
and that it is finite and it remains finite even in the
case with $\langle \xi \rangle = 0$, i.e., $b=a/(1+a)$, and
in particular in the case of a symmetric jump-size distribution,
i.e., $a=1$ and $b=1/2$.
In the limit $s \to 0$ we have
\be
\widetilde{T}'_{<}(s) \sim 
-\frac{1}{s^2[-|v| + \langle \xi \rangle/\langle \tau \rangle]} 
+ \frac{b \, \widetilde{T}'_{<}(a)}{s \langle \tau \rangle 
[-|v| + \langle \xi \rangle/\langle \tau \rangle]} \,,
\quad s \to 0 \,,
\ee
from which, 
by remembering that it holds $T'_<(0)=0$ and that the inverse Laplace
transformation of $1/s$ is a constant, 
%then it results $\widetilde{T}'_{<}(a)=0$, too, 
the asymptotic behaviour is
\be
T'_{<}(x_0) \sim 
- \frac{\langle \tau \rangle} 
{\langle \xi \rangle -|v| \langle \tau \rangle} 
\, x_0 \,,
\quad -|v| \langle \tau \rangle + \langle \xi \rangle < 0 \,,
\quad x_0 \to \infty \,.
\label{Tneg}
\ee

%\textcolor{red}{REMARK by MARCUS: 
%The behaviour $T'_{<}(x_0)$ for small values of $x_0<<1$ can be very different to the asymptotic behaviour of $T'_{<}(x_0)$ for $x_0\to \infty$. It is not said that the asmptotic limit has to be in agreement with the condition $T'_{<}(0)=0$. Our evaluation of the asymptotic limit, $x_0\to\infty$, turns out to be the determination of the leading term only as it can be proven numerically and analytically. For instance the figure in the sent presentation (slide 25) shows that $T'_{<}(0)=0$ is indeed fulfilled by $T'_{<}(x_0)$ but not by its asymptotic behaviour. This may be seen already for one-sided jumps (in the target's direction). For the analytical result please refer both, the last formular on slide 24, last line (exponential waiting times) and slide 25, first line (constant waiting times).}
%
A solution of equation (\ref{MFPT_positive_v}) can indeed be obtained
in the case of the asymmetric exponential PDF (\ref{expPDF})
by assuming
\be
T'_>(x_0) = \sum_{n=0}^\infty a_n x_0^n \,,
\ee
that from (\ref{MFPT_positive_v}) gives
\be
T'_>(x_0) = -
\frac{\langle \tau \rangle}{v \langle \tau \rangle + \langle \xi \rangle} 
\, (1 + x_0) \,, 
\quad v \langle \tau \rangle + \langle \xi \rangle < 0 \,,
\ee
and in the asymptotic and universal limit tends to
\be
T'_>(x_0) \sim 
- \langle \tau \rangle 
\frac{x_0}
{\langle \xi \rangle + v \langle \tau \rangle} \,, 
\quad 
\langle \xi \rangle + v \langle \tau \rangle < 0 \,, 
\quad x_0 \gg 1 \,, %+ v \langle \tau \rangle \,,
\label{Tpos}
\ee
that is consistent with the case $v < 0$ (\ref{Tneg}).

We remark that, in (\ref{Tneg}) and (\ref{Tpos}),
the MFPT depends on the mean value only of the jump-sizes $\langle \xi \rangle$
but, when we move to the dimensional case,
they indeed depend on the whole distribution of the jump-sizes in analogy with
the discussion around formula (\ref{nonuniversal}). Moreover,
also in cases with $\langle \xi \rangle=0$, 
and in particular in the symmetric case, 
the MFPT is finite with an arriving target because 
the condition in (\ref{Tneg}) is always met while
for a departing target the condition in (\ref{Tpos}) 
is never met.

Furthermore, we underline the fact that in the asymptotic limit
for a far-away initial position $x_0 \to +\infty$, 
then a far-away length-scale exists that takes into account the
time-scale of the waiting-times in terms of the mean value such that
the universal behaviour of the MFPT is recovered 
in both reference frames $\mS$ and $\mS'$.
In particular, in $\mS'$ 
the dependence on the whole distribution of the waiting-times 
reduces to the dependence on the mean value only in analogy with $\mS$.

We observe that by comparing formula (\ref{asymptotic}) in $\mS$ and
formulae (\ref{Tneg}) and (\ref{Tpos}) in $\mS'$ the MFPT is
Galilean invariant for two-sided jumps distribution with existing moments
and far-away initial position.
 
If the jump distribution decays with a power-law,
then a finite MFPT is obtained when the condition
$\langle \xi \rangle + v \langle \tau \rangle < 0$ is met.
In particular,
the MFPT is finite in both cases $v < 0$ and $v > 0$ 
whenever the jump-distribution is skewed
in the direction of the target such that $\langle \xi \rangle \to - \infty$. 
This implies that the tail in the direction of the target
decreases according to $|x|^{-(\alpha+1)}$ with $0<\alpha \le 1$
while in the opposite direction it decreases exponentially or
with a power-law $|x|^{-(\alpha'+1)}$ with $1 < \alpha' < 2$.
In analogy with the one-sided case studied in the next subsection,
we have that for $x_0 \to + \infty$ 
the MFPT goes like $x_0^\alpha$ with $0 < \alpha \le 1$ 
(where the extremal value $\alpha=1$ is included 
for taking into account when the jumps
in the direction of the target are distributed according to the left-side
of the Cauchy distribution, which leads to an infinite negative mean value).
A noteworthy two-sided special case is provided by symmetric 
L\'evy densities with $1 < \alpha < 2$ when $v < 0$. 
In fact, in this case we have $\langle \xi \rangle =0$ and
a finite MFPT proportional to $x_0$ follows
from the uniform motion of the target.

\subsection{One-sided jumps}
In order to guarantee a finite MFPT
also in the case with power-law jump distribution,
in analogy with the analysis in $\mS$, see formula (\ref{Teq}), we assume 
one-sided jumps and the equation (\ref{eq18}) for $T'(x_0)$ becomes 
\begin{eqnarray}
T'(x_0)&=& 
\int_0^\infty \!\! \Psi\lk \tau\rk \Theta\lk v\tau + x_0 \rk d\tau 
\nonumber \\
& & 
+ \int_0^\infty \!\! \psi\lk \tau \rk \Theta\lk v\tau + x_0 \rk 
\int_0^{x_0+v\tau} \!\! q\lk \xi-x_0-v\tau \rk T'\lk \xi \rk \, d\xi d\tau \,, 
\end{eqnarray}
%or, by highlighting the sign of the velocity $v$,
%we have
%\begin{subnumcases}{\hspace{-1.0truecm}T'(x_0)=}
%\displaystyle{
%\int_0^{-x_0/v} \Psi\lk \tau\rk d\tau 
%+ \int_0^{-x_0/v} \!\! \psi\lk \tau \rk \int_0^{x_0+v\tau} 
%\!\!\!\! 
%q\lk \xi-x_0-\eta\tau \rk T'\lk \xi \rk d\xi  d\tau \,, \quad v<0 \,,} \\
%\nonumber \\
%\displaystyle{
%\langle \tau \rangle + \int_0^\infty \!\! \psi\lk \tau \rk 
%\int_0^{x_0+v\tau} \!\!q\lk \xi-x_0-v\tau \rk T'\lk \xi \rk d\xi d\tau \,,
%\quad v\ge 0 \,.}
%\label{eq20}
%\end{subnumcases}
%Therefore, from the comparison against (\ref{Teq}),
%we observe that the MFPT estimated by an observer in $\mS'$ 
%requires the knowledge of the whole waiting-times PDF $\psi(\tau)$ in 
%opposition to the estimation by an observer in $\mS$.
%
%We denote 
%\begin{subnumcases}{T'(x_0)=}
%\displaystyle{
%T'_<(x_0; \sigma) \,, \quad -\sigma = v<0 \,,} \\
%\nonumber \\
%\displaystyle{
%T'_>(x_0; \omega) \,, \quad \omega = v \ge 0 \,,}
%\end{subnumcases}
such that formulae (\ref{MFPT_negative_v}) and (\ref{MFPT_positive_v}) 
now read 
\be
T'_<\lk x_0\rk = 
\int_0^{x_0/|v|} \hspace{-0.5truecm} \Psi\lk \tau\rk d\tau + 
\int_0^{x_0/|v|} \hspace{-0.5truecm} \psi\lk \tau \rk \int_0^{x_0-|v|\tau} 
\hspace{-0.5truecm} q\lk \xi-x_0+|v|\tau \rk T'_< (\xi) d\xi d\tau \,,
\, v<0 \,, 
\label{eq21a} 
\ee
\be
T'_>\lk x_0 \rk = \langle \tau \rangle + 
\int_0^\infty \psi\lk \tau \rk \int_0^{x_0+v\tau} 
q\lk \xi-x_0-v\tau \rk T'_> \lk \xi \rk d\xi d\tau \,,
\quad v\ge 0 \,.
\label{eq21}
\ee
Therefore, from the comparison against (\ref{Teq}),
we observe that the MFPT estimated by an observer in $\mS'$ 
requires the knowledge of the whole waiting-times PDF $\psi(\tau)$ in 
opposition to the estimation by an observer in $\mS$.
In particular,
\be
T'(x_0)=
T'_< \lk x_0 \rk = T'_> \lk x_0 \rk 
= \langle \tau \rangle + 
\int_0^{x_0} 
q\lk \xi-x_0 \rk T' \lk \xi \rk d\xi d\tau \,,\quad {\rm for }\quad v=0\,,
\label{Tprimeeq}
\ee
that is equal to the corresponding formula (\ref{Teq})
in $\mS$, when $x_0 > 0$, as it is expected.

\subsubsection{One-sided jumps with arrival target: $v < 0$}
\label{sec:posoneside}
In the case of an arrival target, namely the case when $v<0$,
we can derive an explicit solution of $T_< \lk x_0\rk$ 
from formula (\ref{eq21a}) for a one-sided jump PDF in analogy
with (\ref{eq11}).
In fact, by applying Laplace transformation to (\ref{eq21a}) 
we have
\begin{eqnarray}
\widetilde{T}'_< (s)
&=&\frac{\widetilde\Psi\lk s|v|\rk}
{s \, \left[1-\widetilde\psi( s|v|) \widetilde{\rho}(s)\right]} \nonumber \\
&=&\frac{\widetilde\Psi\lk s|v|\rk}{s} 
\sum_{n=0}^\infty \left[
\widetilde{\psi} \lk s |v|\rk \widetilde{\rho}(s) \right]^n \nonumber \\
&=&\frac{1}{s} \sum_{n=0}^\infty 
\widetilde{\Psi}_n \lk s|v| \rk \widetilde{\rho}_n(s) \,,
\label{eq22}
\end{eqnarray}
and then
\be
T'_< (x_0)
=\frac{1}{|v|} \sum_{n=0}^\infty
\int_0^{x_0} \int_0^{x_0-z} 
\Psi_n\lk\frac{y}{|v|}\rk \rho_n(x_0-z-y) \, dydz \,,
\label{eq22a}
\ee
where $\rho(-\xi) = q(\xi)$ and $\rho_n(\xi)$ are 
the same as in (\ref{eq11}), and
$\Psi_n(\tau)$ is the probability of $n$ jumps in $[0,\tau]$,
i.e., $\displaystyle{
\Psi_n(\tau)=\int_0^\tau \Psi_{n-1}(\tau-\chi)\psi(\chi)d\chi}$ 
with $\Psi_0(\tau)=\Psi(\tau)$.

By comparing (\ref{eq22a}) and (\ref{eq11}),
we obtain that the MFPT in the reference frame $\mS'$ 
depends on the whole waiting-times PDF $\psi(\tau)$,
more precisely on the probability $\Psi(\tau)$,
in opposition to the MFPT in the reference frame $\mS$ that
depends on the mean value only of the waiting-times.
We stress that when the target is arriving,
i.e., $v < 0$,
there is not any special case of the jump PDF
such that the MFPT can be determined by the mean waiting-times only 
rather than by the whole PDF $\psi(\tau)$ in analogy with the cases
in $\mS$ and, as we will see in the next section, 
in the case with departing target, i.e., $v >0$,
and one-sided jumps.

If we consider a jump PDF with existing moments such that in the 
limit $s \to 0$ it holds $\widetilde{\rho}(s) \sim 1 + \langle \xi \rangle s$,
with $\langle \xi \rangle < 0$, then from the first line of (\ref{eq22}) 
we obtain
\be
T'_< (x_0) \sim - 
\frac{\langle \tau \rangle}{\langle \xi \rangle - |v| \langle \tau \rangle} 
\, x_0\,,
\quad x_0 \to \infty \,.
\label{Tnegoneside}
\ee 

If, if in analogy with \S \ref{onesideS}, 
we consider also a one-sided L\'evy distribution
for the jumps then $\widetilde{\rho}(s)=\e^{-s^\alpha}$, 
with $0 < \alpha < 1$, then from the first line of (\ref{eq22}) it holds
\be
\displaystyle{
\widetilde{T}'_<(s) = 
\frac{\widetilde{\Psi}(|v| s)}
{s \, [1-\widetilde{\psi}(|v| s) \, \e^{-s^\alpha}]} \,,
}
\label{Toneside}
\ee
that in limit $s \to 0$ gives
\be
\widetilde{T}'_<(s) \sim \frac{\langle \tau \rangle}{s^{1+\alpha}} \,,
\quad s \to 0 \,,
\ee
where we have used the fact that 
$\widetilde{\Psi}(|v| s)=(1-\widetilde{\psi}(|v| s))/(|v| s)$ and
the limit behaviour
$\widetilde{\psi}(|v| s) \sim 1 - |v| \langle \tau \rangle s$
for $s \to 0$
such that in the asymptotic case $x_0 \to +\infty$ it results
\be
T'_<(x_0) \sim \frac{\langle \tau \rangle}{\Gamma(1+\alpha)} \, x_0^\alpha \,,
\quad 0 < \alpha < 1 \,,
\quad x_0 \to +\infty \,,
\label{MFPTonesideSprime}
\ee
that is equal to formula (\ref{MFPTonesideS}) in $\mS$ for
the similar case of one-sided jumps in the direction of the absorbing boundary.

By comparing (\ref{Tnegoneside}) against (\ref{TonesideS}) and 
(\ref{MFPTonesideSprime}) against (\ref{MFPTonesideS}) 
it follows that, when the target is arriving, i.e., $v < 0$, 
the MFPT is a Galilean invariant for one-sided jumps
with existing moments and such invariance is indeed not met
for one-sided jumps with power-law distribution. 

\subsubsection{One-sided jumps with departing target: $v > 0$}
In the case of a departing target, namely the case when $v > 0$,
we can not use the technique of the Laplace transform.
Therefore, first set formula (\ref{eq21}) for a one-sided jump PDF in analogy
with (\ref{eq11}), i.e., 
\be
T'_>\lk x_0 \rk
=\langle \tau \rangle + \int_0^\infty \psi\lk \tau \rk 
\int_0^{x_0+v\tau} \rho\lk x_0+v\tau-\xi \rk 
T'_>\lk \xi \rk \, d\xi d\tau \,,
\ee 
and later we split the problem as follows
\begin{subnumcases}{}
T'_> \lk x_0\rk=
\langle \tau \rangle + 
\int_0^\infty \psi\lk \tau \rk h(x_0+v\tau) \, d\tau \,, 
\label{eq22bup} \\
\nonumber \\
h(y)=\int_0^{y} \rho\lk y-\xi \rk T'_>\lk \xi \rk \, d\xi \,.
\label{eq22b}
\end{subnumcases}
In this case we can not derive the explicit form of $T'_>(x_0)$
in analogy with (\ref{eq11}) and (\ref{eq22a}). 
But, we can study when the MFPT $T'_>(x_0)$ depends
on the mean value only of the waiting-times and not on the
whole PDF $\psi(\tau)$, see Appendix \ref{AppendixB} for the detailed 
study. As we have already seen, this is analog to study the asymptotic
behaviour when $x_0 \to +\infty$.
From (\ref{eq22bup}), we observe that  
$T'_>\lk x_0 \rk$ depends 
on the mean value only of the waiting-times if 
\be
h(y)=a_3 y + a_2 \,, 
\label{eq22c}
\ee
where $a_3$ and $a_2$ are constants to be determined
(see, Appendix \ref{AppendixB})
and from definition (\ref{eq22b}) we have that $h(y) \ge 0$ with $y \ge 0$. 
Moreover, 
if we disregard the possibility for the walker to stay in the actual position 
we have that $h(0)=0$ 
and therefore it follows that $a_3 > 0$ and $a_2=0$.
Otherwise, if the walker may stay in the actual position with
non null probability then 
$a_2 = - c/(\langle \xi \rangle/\langle \tau \rangle + v) \ne 0$ 
(see notation and formula (\ref{parameter_2}) in Appendix \ref{AppendixB}).

By plugging (\ref{eq22c}) into (\ref{eq22bup}) then
\be
T'_>\lk x_0\rk = 
a_3 x_0 + (1+a_3 v) \langle\tau\rangle \,,
\label{Tposoneside0}
\ee
and the Laplace transform reads
\be
\widetilde{T}'_> \lk s \rk 
= \frac{a_3 + (1+a_3 v)\langle\tau\rangle \, s}{s^2} \,.
\label{TLaplace}
\ee
To conclude, from equalling the Laplace transforms of
formulae (\ref{eq22b}) and (\ref{eq22c}) and by using (\ref{TLaplace})
we have that $\widetilde{\rho}(s)$ emerges to be
\be
\widetilde{\rho}(s) = 
\frac{a_3}{a_3 + \lk 1 + a_3 v\rk \langle\tau\rangle \, s} 
\,.
\label{eq22d}
\ee 
By remembering the setting stated in (\ref{theta}),
parameter $a_3$ is determined by fixing a finite-mean jump PDF 
through the formula
$\displaystyle{d\widetilde{\rho}(s)/ds|_{s=0}=
-\langle \theta \rangle=\langle \xi \rangle < 0}$ 
which provides
\be
a_3 = - \frac{\langle\tau\rangle}
{\langle \xi \rangle + v  \langle\tau \rangle} > 0 \,,
\quad \langle \xi \rangle + v \langle \tau \rangle < 0 \,,
\quad \langle \xi \rangle < 0 \,,
\label{eq22e}
\ee
that inserted into (\ref{Tposoneside0}) gives
\be
T'_>(x_0) \sim - 
\frac{\langle \tau \rangle}
{\langle \xi \rangle + v \langle \tau \rangle} \, x_0 \,,
\quad x_0 \gg - \langle \xi \rangle \,.
\label{Tposoneside}
\ee
On the other side, if we insert (\ref{eq22e}) into (\ref{eq22d}) 
we obtain
\be
\widetilde{\rho}(s)
=\frac{1}{1 - \langle \xi \rangle \, s} \,,
\label{eq22f}
\ee
and, by remembering the notation setting (\ref{theta})
with $\theta=-\xi \in \RR^+_0$ such that 
$\langle \theta \rangle= - \langle \xi \rangle$,
the jump PDF takes the exponential form
\be
\rho(\theta) 
= \frac{1}{\langle \theta \rangle} \, \e^{-\theta/\langle \theta\rangle} 
= - \frac{1}{\langle \xi \rangle} \, \e^{-\xi/\langle \xi \rangle} 
= q(\xi) \,, 
\quad \theta \in \RR^+_0 \,,
\quad \xi \in \RR^-_0 \,.
\label{eq22g}
\ee 
Hence, an exponential distribution of the jump-lengths 
allows for an estimation of the MFPT that is 
dependent on the mean-value only of the waiting-times, 
and it is independent of the whole distribution $\psi(\tau)$,
in analogy with the estimations in the reference frame $\mS$. 

In the case of a one-sided power-law distribution of jump-sizes,
i.e., $\widetilde{\rho}(s)= \e^{-s^\alpha}$,
the estimation of the MFPT can be derived as follows.
Motivated by the case analysed in \S \ref{sec:posoneside}, 
we assume that $T'_>(x_0) \sim A_1 \, x_0^\alpha$, with 
$A_1 \propto \langle \tau \rangle >0$,
and then from the Laplace transform of (\ref{eq22b}) in the limit
$s \to 0$ we have
\be
\widetilde{h}(s) \sim A_1 \, \frac{1 - A_0 \, s^\alpha}{s^{\alpha+1}} \,,
\quad s \to 0 \,,
\label{htilde}
\ee
where $A_0 > 0$ is a constant to be determined.
From (\ref{htilde}) we have that 
\be
h(y) \sim A_1 \, y^\alpha - A_1 A_0 \,, \quad y \to + \infty \,,
\ee
and if this is plugged in (\ref{eq22bup}) we obtain
\be
T'_>(x_0) \sim \langle \tau \rangle 
+ A_1 x_0^\alpha 
\int_0^\infty \psi(\tau) \left[1 + \frac{v \tau}{x_0}\right]^\alpha 
\, d\tau 
- A_1 A_0
\,. 
\ee
We introduce now the parameter $\tau_*$ such that 
$v \tau/x_0 \ll 1$ when $\tau < \tau_*$ and thus
\begin{eqnarray}
T'_>(x_0) &\sim& \langle \tau \rangle 
+ A_1 x_0^\alpha 
\int_0^{\tau_*} \psi(\tau) \left[1 + \alpha \frac{v \tau}{x_0}\right] 
\, d\tau \nonumber \\
& & \qquad \qquad
+ A_1 x_0^\alpha 
\int_{\tau_*}^\infty \psi(\tau) \left[1 + \frac{v \tau}{x_0}\right]^\alpha 
\, d\tau 
- A_1 A_0
\,. 
\end{eqnarray}
Since $\tau_* \to +\infty$ when $x_0 \to +\infty$,
then the contribution of the second integral can be neglected 
and we finally get
\be
T'_>(x_0) \sim \langle \tau \rangle 
+ A_1 B_1 \, x_0^\alpha 
+ A_1 B_0 \alpha v \, x_0^{\alpha-1} - A_1 A_0
\,,
\ee
where $\displaystyle{
B_1=\int_0^{\tau_*}\psi(\tau)d\tau \sim \mathcal{O}(1)}$ and 
$\displaystyle{B_0=\int_0^{\tau_*}\tau\psi(\tau)d\tau \sim \mathcal{O}(
\langle \tau \rangle)}$.
To conclude,
by setting $\langle \tau \rangle \propto A_1 A_0$ we have 
$A_0 \sim \mathcal{O}(1)$ such that
\be
T'_>(x_0) \sim \langle \tau \rangle \, x_0^\alpha \,, 
\quad x_0 \to +\infty \,,  
\label{TposonesideLevy}
\ee
that was the assumed estimation of the MFPT.

By comparing (\ref{Tposoneside}) against (\ref{TonesideS}) and 
(\ref{TposonesideLevy}) against (\ref{MFPTonesideS}) 
it follows that, when the target is departing, i.e., $v > 0$, 
the MFPT is a Galilean invariant for one-sided jumps
with existing moments but such invariance is indeed broken
for one-sided jumps with a power-law distribution. 

\section{Summary and Conclusion}
\label{conclusions}
In the view of the failure of present
diffusive models regarding the Galilean invariance 
\cite{cairoli_etal-pnas-2018},
we have analysed such failure with respect to the MFPT
in the case of the CTRW approach 
\cite{kutner_etal-epjb-2017,shlesinger-epjb-2017},
that
is one of the most suitable stochastic processes for 
studying diffusion in biological systems 
\cite{barkai_etal-pt-2012,goiko_etal-bj-2018}
where the MFPT finds one of the most important applications 
\cite{metzler_etal-firstpassagetime-book-2014,
jpaspecialissue-2020,bressloff_etal-2014,
chou_etal-2014,polizzi_etal-ijc-2016}.

The main result is that, for an observer co-moving with the target, 
the MFPT does not depend on the specific waiting-times distribution
except for its mean value while, 
for an observer in uniform motion with respect the target,
the specific distribution of the waiting-times plays a role.
However, there exists a length-scale determined by the
time-scale as well that defines a far-away initial position such
that the universal behaviour is observed in both reference frames
and the Galilean invariance is met for the MFPT when it exists.
The estimations in the two reference frames are reconciled
with respect to the independence of the whole waiting-times 
distribution, only when the relative velocity between 
the target and the observer is positive and 
the (two-sided) jump-distribution in the moving reference frame is 
of a special form combining a one-sided exponential 
and a delta-distribution
together with an arbitrary jump-size distribution with finite mean
in the opposite direction with respect to the target
(see the derivation in Appendix \ref{AppendixB}, 
in particular the general definition of the jump-distribution 
\eqref{distinction_jumps}, its setting 
in \eqref{jump_average_neg} and \eqref{jump_average_pos} 
and finally formula \eqref{eq22k}).

Summarising our results we found that:

In the reference frame $\mS$ co-moving with a target, 
the MFPT depends on the jump PDF and on the mean value only
of the waiting-times: 
this occurs with both two-side and one-side jumps' distribution. 
When a jump PDF with one-sided jumps in the direction of the target 
is considered, the MFPT emerges to be finite also in the case of a 
jump PDF with a power-law tail, namely a one-side L\'evy density, 
and it grows with a sublinear law in
opposition to the linear law of the case with an exponential jump PDF.
When the two-sided jump PDF decreases with exponential tails
then the MFPT is finite if the finite mean jump-size 
is negative, as well as in the case with power-law tails for 
which the mean jump-size is negative but infinite.
However also in this case the MFPT can turn into infinite
if the mean jump-size is non-negative, as it occur with
an asymmetric or symmetric jump PDF.

In the reference frame $\mS'$, that moves with a uniform motion with 
respect to the target through a Galilean transformation, 
the MFPT depends on both the PDFs of jump-sizes and waiting-times
and in both two-sided and one-sided case. 
Nevertheless, for initial positions sufficiently far-away from the 
target, the dependence on the exact waiting-times distribution 
disappears and only the dependence on the mean waiting-time remains.
For an arriving target,
the MFPT is finite even when the mean jump-size is zero, 
which includes the case of a symmetric jump distribution, 
because condition \eqref{Tneg} is always met.
On the contrary, for a departing target the corresponding condition \eqref{Tpos} 
is never met when the mean jump-size is zero. 
For an observer that is moving in both directions with respect to the target,
it emerges that the MFPT is finite 
only in the case of an asymmetric jumps-size distribution with finite mean 
and if the averaged relative shift between the observer 
in $\mS'$ and the target is negative. 
In this setting the MFPT meets Galilean invariance.
In the case of power-law tails, the MFPT is finite
whenever the jump-distribution is skewed: which implies
a condition on the decaying of the tail in the direction of the
target and also in the opposite direction. 
However, a noteworthy special case with finite MFPT emerges 
when a symmetric jump distribution with $1 < \alpha < 2$
is consider with arrival target, 
because in this setting the MFPT is determined by the uniform 
motion of the reference frame.
Moreover, we report that in the one-sided case with 
power-law distributed jumps in the direction of the target, 
the MFPT does not differ from the MFPT for a co-moving 
observer that is characterised through the same power-law distribution of jumps 
towards the target. 
This fact highlights that the MFPT of a CTRW model is not
a Galilean invariant.

In a long perspective, 
under the assumption that the CTRW is the proper approach
for modelling a phenomenon, see, e.g.,
\cite{zhang2020identification,shen2021kinetics,
awad2021continuous,levin2021measurements}, 
this theoretical result can be understood as a method for an 
indirect measurement of the waiting-times distribution,
in opposition to existing studies on the MFPT of a CTRW
in a co-moving reference frame $\mS$, 
see, e.g., \cite{dahlenburg_etal-jpa-2022}, 
that allow indeed only for the measurement of the mean waiting-time.

\subsection*{Acknowledgments}

This research is supported by the Basque Government through the 
BERC 2022--2025 program, 
by the Ministry of Science and Innovation: 
BCAM Severo Ochoa accreditation 
CEX2021-001142-S / MICIN / AEI / 10.13039/501100011033, and through the Predoc Severo Ochoa 2018 grant PRE2018-084427.

\appendix

\setcounter{equation}{0}
\renewcommand{\theequation}{{\thesection}.\arabic{equation}}
\section{Finite MFPT for L\'evy flights}
\label{AppendixA}
The MFPT for L\'evy flights is finite whenever it holds
\cite[{\it mutatis mutandis} from formulae (46) and (47)]
{padash_etal-jpa-2019}
\be
\frac{\alpha\pi}{2} <
\arctan\left\{\beta\tan \frac{\alpha\pi}{2}\right\} \,,
\quad \alpha \in (0,1) \cup (1,2) \,,
\quad \beta \in (-1,1) \,.
\label{condition}
\ee
When $0 < \alpha < 1$ then $\tan \alpha\pi/2$ is a 
positive monotonically increasing function,
therefore condition (\ref{condition}) turns into 
\be
\tan \frac{\alpha\pi}{2} <
 \beta \tan \frac{\alpha\pi}{2} \,,
\ee
that is never met.
When $1 < \alpha < 2$ then $\tan \alpha\pi/2$ is a 
negative monotonically increasing function,
therefore condition (\ref{condition}) turns into 
\be
- \left\vert \tan \frac{\alpha\pi}{2} \right\vert <
- \beta \left\vert \tan \frac{\alpha\pi}{2} \right\vert \,,
\ee
that is always met. To conclude, the MFPT for asymmetric L\'evy densities
is finite if $1 < \alpha < 2$.

%It is known that if the jump PDF is a stable L\'evy density
%then the walker distribution is a L\'evy stable, too, 
%see, e.g., \cite{zaburdaev_etal-rmp-2015}, 
%and the MFPT can be estimated by the integral
%\cite[formula (28)]{padash_etal-jpa-2020}
%\be
%T(x_0)=\int_0^\infty \left\{
%\int_0^\infty 
%t^{-1/\alpha} \mathcal{L}_\alpha^\theta(t^{-1/\alpha}(x-x_0)) \, dx
%\right\} dt \,,
%\label{MFPTLevy0}
%\ee
%where $\mathcal{L}_\alpha^\theta(y)$ is a L\'evy stable density
%of stability index $\alpha \in (0,2)$ and
%asymmetry parameter $|\theta| \le \min\{\alpha,2-\alpha\}$
%\cite{mainardi_etal-fcaa-2001}. 
%An alternative procedure for deriving the existing literature results
%is the following. By remembering that it holds 
%\cite[{\it mutatis mutandis} in formula (6.14)]{mainardi_etal-fcaa-2001}
%\be
%\int_0^\infty 
%\mathcal{L}_\alpha^\theta(y) \, dy = \frac{\alpha-\theta}{2 \, \alpha} \,,
%\ee
%by applying the change of variable $t^{-1/\alpha}(x-x_0)=y$
%from formula (\ref{MFPTLevy0}) we obtain
%\be
%T(x_0)=
%\int_0^\infty \left\{
%\int_0^{x_0 t^{-1/\alpha}} 
%\mathcal{L}_\alpha^\theta(-y) \, dy \right\} dt +
%\frac{\alpha-\theta}{2 \, \alpha} 
%\int_0^\infty dt \,,
%\label{MFPTLevy1}
%\ee
%which diverges except when $\theta=\alpha$ when $\alpha \in (0,1)$.
%We remind that $\mathcal{L}_\alpha^\alpha(y)$, 
%with $\alpha \in (0,1)$, is a one-sided density with support $\RR_0^-$
%and that in the limit $\alpha \to 1$ it holds $\mathcal{L}_1^1(y)=\delta(y+1)$.%The one-sided case is considered separately.

\setcounter{equation}{0}
\renewcommand{\theequation}{{\thesection}.\arabic{equation}}
\section{When the MFPT in $\mS'$ depends
on the mean waiting time only}
\label{AppendixB}
We derive here the most general jump-distribution for which 
$T'_{>}\lk x_0\rk$ depends on the mean waiting time only.
We start from (\ref{MFPT_positive_v}) that we rewrite as
\begin{equation}
\label{MFPT_positive_v_rho}
T'_{>}\lk x_0 \rk = \langle\tau\rangle + 
\int_0^{\infty} \psi\lk \tau \rk 
\int_0^{\infty} \rho\lk x_0+v\tau-\xi\rk T'_{>}(\xi) \, d\xi d\tau \,,
\quad v>0 \,,
\end{equation}
where 
\begin{eqnarray}
\label{distinction_jumps}
\rho(\xi)=q(-\xi)=
\left\{
\begin{array}{l l l}
(1-b)\,\rho_>(\xi),&\text{ if }\xi\in \mathbb{R}_0^+ ,\\
\\
b\,\rho_<(\xi),&\text{ if }\xi\in\mathbb{R}^- ,
\end{array}
\right.\ 
\end{eqnarray}
without any further constraint than the normalisations
$\displaystyle{\int_0^\infty \rho_>(\xi)\,d\xi=1}$ 
and $\displaystyle{\int_{-\infty}^0 \rho_<(\xi)\,d\xi=1}$ 
such that $b$ is the probability to jump in the opposite direction
to the target.
Moreover we define
\begin{eqnarray}
\int_0^\infty \xi\,\rho_>(\xi)\,d\xi
= -\langle\xi\rangle^{-} \,, \quad\langle\xi\rangle^{-}\in\mathbb{R}_0^- \,,
\label{jump_average_neg}\\
\int_{-\infty}^0 \xi\,\rho_<(\xi)\,d\xi
= -\langle\xi\rangle^{+} \,, \quad\langle\xi\rangle^{+}\in\mathbb{R}^+ \,,
\label{jump_average_pos}
\end{eqnarray}
such that 
$\displaystyle{
\langle\xi\rangle
= \int_\mathbb{R} \xi q(\xi) d\xi
= -\int_\mathbb{R} \xi \rho(\xi) d\xi
= (1-b) \langle\xi\rangle^{-}+b \langle\xi\rangle^{+}}$.
By using \eqref{distinction_jumps}, 
equation \eqref{MFPT_positive_v_rho} can be rewritten as
\begin{eqnarray}\label{MFPT_positive_v_rho_2}
T'_{>}\lk x_0 \rk
&=& \langle\tau\rangle 
+ \int_0^{\infty} \psi\lk \tau \rk  
\lkk (1-b)\int_0^{x_0+v\tau} \rho_>\lk x_0+v\tau-\xi\rk T'_{>}(\xi) \, d\xi 
\right. \nonumber \\
& & \qquad \qquad  
\left. +b\int_{x_0+v\tau}^{\infty} \rho_<\lk x_0+v\tau-\xi\rk T'_{>}(\xi)
\, d\xi\rkk d\tau \nonumber \\
&=& \langle\tau\rangle 
+ \int_0^{\infty} \psi\lk \tau \rk (1-b) h(x_0+v\tau) \, d\tau \nonumber \\
& & \qquad 
+ \, b \int_{-\infty}^{0} \rho_<\lk \xi\rk \int_0^{\infty} 
\psi\lk \tau \rk T'_{>}(x_0+v\tau-\xi) \, d\tau d\xi \,,
\end{eqnarray}
where
\begin{equation}
\label{define_h}
h(y)=\int_0^y \rho_> \lk y-\xi\rk T'_{>}(\xi) \, d\xi \,.
\end{equation}
From \eqref{MFPT_positive_v_rho_2}, we have that $T'_{>}\lk x_0\rk$ depends
on the mean value of the waiting-times only if
\begin{equation}\label{lin_app}
\left\{
\begin{array}{l  l}
T'_{>}\lk x_0 \rk = a_1 \, x_0+a_0 \,,\\
\\
h(y)=a_3 \, y+a_2 \,.
\end{array}
\right.\ 
\end{equation}
By plugging \eqref{lin_app} into \eqref{MFPT_positive_v_rho} we obtain
\begin{eqnarray}\label{MFPT_lin_app}
a_1 \, x_0+a_0
&=& \langle\tau\rangle 
+ \int_0^{\infty} \psi\lk \tau \rk (1-b)
\lk a_3 \,(x_0+v\tau)+ a_2\rk \, d\tau \nonumber \\
& & 
+ b\,\int_{-\infty}^{0} \rho_<\lk \xi\rk 
\int_0^{\infty} \psi\lk \tau \rk \lk a_1 \, (x_0+v\tau-\xi)+a_0\rk \, 
d\tau\, d\xi \nonumber \\
&=& \langle\tau\rangle 
+ (1-b)\lk a_3\,(x_0+v\langle\tau\rangle) + a_2\rk + \nonumber \\
& & b \,\int_{-\infty}^{0} \rho_<\lk \xi\rk \lk 
a_1 \, (x_0+v\langle\tau\rangle-\xi)+a_0\rk\, d\xi \nonumber \\
&=& \langle\tau\rangle 
+ (1-b)\lk a_3\,(x_0+v\langle\tau\rangle) + a_2\rk + \nonumber \\
& & 
b \, \lk a_1\, (x_0+v\langle\tau\rangle+\langle\xi\rangle^+)+a_0\rk 
\nonumber \\
&=& \lk b\,a_1+(1-b)\,a_3\rk\,x_0 + \langle\tau\rangle 
+ (1-b)\lk a_3\,v\langle\tau\rangle+a_2\rk \nonumber \\
& &  
+b\, \lk a_1\, (v\langle\tau\rangle+\langle\xi\rangle^+)
+ a_0\rk \,.
\end{eqnarray}
Comparing the coefficients of monomials of $x_0$ gives
\begin{eqnarray}
a_1 &=& a_3 \,,\label{coeff_compar_a1}\\
a_0 &=& \frac{\langle\tau\rangle\lk 1 + a_3 \, v\rk+(1-b)\,a_2 
+ b\, a_3 \,\langle\xi\rangle^+}{1-b} \,,
\label{coeff_compar_a0}
\end{eqnarray}
and it holds
\be
T'_{>}\lk x_0 \rk
= a_3\,x_0 + \frac{\langle\tau\rangle\lk 1 
+ a_3\,v\rk+(1-b)\,a_2 +b\, a_3\,\langle\xi\rangle^+}{1-b} \,,
\label{MPFT_lin_app_2}
\ee
and in the Laplace domain
\be
\widetilde{T}'_{>}\lk s \rk
= \frac{a_3}{s^2} + \frac{\langle\tau\rangle\lk 1 
+ a_3\,v\rk+(1-b) \, a_2 + b\, a_3 \,\langle\xi\rangle^+}
{(1-b)s} \,.
\label{MPFT_lin_app_2_lap}
\ee
The Laplace transform of formula \eqref{define_h} 
together with \eqref{MPFT_lin_app_2_lap} provide 
the Laplace transform of $\rho_>(\xi)$, that is
\be
\label{h_laplace}
\widetilde{\rho}_>(s)
= \left\{1+s\,\lk\frac{\langle\tau\rangle\lk 1 
+  a_3\,v\rk +b\, a_3\,\langle\xi\rangle^+}{(1-b)(a_3+a_2\, s)}\rk
\right\}^{-1} \,.
\ee
Coefficient $a_3$ can be determined by noting from 
\eqref{jump_average_neg} that
\be
\label{rho_aver_lap}
\langle\xi\rangle^-=\frac{d\tilde{\rho}_>(s)}{ds}\bigg{|}_{s=0}
=\frac{-\langle\tau\rangle(1+a_3\,v)
- b\,a_3\,\langle\xi\rangle^+}{a_3\,(1-b)} \,,
\ee
and then
\be
a_3 = - \frac{\langle\tau\rangle}{\langle\xi\rangle+v\langle\tau\rangle}
\label{a3} \,.
\ee
Furthermore, after inserting \eqref{a3} into \eqref{h_laplace} and
by remembering that the jump-PDF $q(\xi)$ and, thus, 
$\rho_>(\xi)$ are independent of $v$ and $\tau$ by definition, 
we have also the coefficient $a_2$: 
\begin{equation}
\label{parameter_2}
a_2 = - c \, \frac{\langle\tau\rangle}{\langle\xi\rangle + v \langle\tau\rangle}
= c \, a_3 \,,
\quad c \in \mathbb{R} \,.
\end{equation}
Finally, we have
\be
\widetilde \rho_>(s)
= \frac{1+sc}{1+s(c-\langle \xi\rangle^-)}
=\frac{-\langle \xi\rangle^-}{(c-\langle \xi\rangle^-)
(1+s(c- \langle \xi\rangle^-))} 
+ \frac{c}{c-\langle \xi\rangle^-} \,,
\ee
and after the inversion
\be
\rho_>(\xi)
=\frac{{-\langle \xi \rangle^-}}{(c-{\langle \xi \rangle^-})^2} \,
\exp\lk\frac{-\xi}{c-\langle \xi\rangle^-}\rk +
\frac{c}{c-{\langle \xi \rangle^-}} \,\delta(\xi) \,,
\quad c\,,\xi\in\mathbb{R}_0^+ \,. 
\label{eq22k}
\ee 
In formula \eqref{eq22k} the weight coefficient 
$c/(c-{\langle \xi \rangle^-})$ 
can be understood as the probability to stay and 
the weight coefficient 
$-\langle \xi \rangle^-/(c - {\langle \xi \rangle^-})$ 
as the probability to jump according to a one-sided exponential distribution 
in the direction of the absorbing boundary. 
Within this interpretation, each weight belongs to the interval $[0,1]$
and then, since $\langle \xi \rangle^- \le 0$, we have that $c \ge 0$.
To conclude, by plugging \eqref{a3} into \eqref{coeff_compar_a0} 
we can determine the coefficient $a_0$ as
\be
a_0 = a_3 (c - \langle\xi\rangle^-) \,,
\label{parameter_1_0}
\ee
and finally the MFPT in \eqref{MPFT_lin_app_2} results to be
\begin{equation}
T'_>(x_0)
= -\frac{\langle \tau \rangle}{\langle \xi \rangle+v\langle \tau \rangle}\,x_0
+ \lk \frac{\langle \xi \rangle^- -c}
{\langle \xi \rangle+v\langle \tau \rangle} \rk \, \langle \tau \rangle\,,
\quad x_0\in\mathbb{R}_0^+ \,.
\label{MFPT_final}
\end{equation}

\end{document}